\documentclass[aps,pra,twocolumn,groupedaddress,superscriptaddress,showpacs]{revtex4}
\usepackage{natbib}
\usepackage[latin1]{inputenc}
\usepackage[dvips]{graphicx}
\usepackage{amsmath,amsbsy,amsfonts,amssymb}
\usepackage{bm}
\usepackage{textcomp}

\begin{document}
\title{Fourier-transform spectroscopy of Sr$_2$ and revised ground state potential}
\author{A. Stein}
\affiliation{Institut f\"ur Quantenoptik, Leibniz Universit\"at Hannover,
Welfengarten 1, D-30167 Hannover, Germany}
\author{H. Kn\"ockel}
\affiliation{Institut f\"ur Quantenoptik, Leibniz Universit\"at Hannover,
Welfengarten 1, D-30167 Hannover, Germany}
\author{E. Tiemann}
\affiliation{Institut f\"ur Quantenoptik, Leibniz Universit\"at Hannover,
Welfengarten 1, D-30167 Hannover, Germany}
\date{\today}
\begin{abstract}
Precise potentials for the ground state X$^1\Sigma^+_g$ and the minimum region of the excited state 2$^1\Sigma^+_u$ of Sr$_2$ are derived by high resolution Fourier-transform spectroscopy of fluorescence progressions from single frequency laser excitation of Sr$_2$ produced in a heat pipe at 950 \textcelsius. A change of the rotational assignment by four units compared to an earlier work (G. Gerber, R. M\"oller, and H. Schneider, J. Chem. Phys. 81, 1538 (1984)) is needed for a consistent description leading to a significant shift of the potentials towards longer inter atomic distances. The huge amount of ground state data derived for the three different isotopomers $^{88}$Sr$_2$, $^{86}$Sr$^{88}$Sr and $^{87}$Sr$^{88}$Sr (almost 60\% of all excisting bound rovibrational ground state levels for the isotopomer $^{88}$Sr$_2$) fixes this assignment undoubtedly. The presented ground state potential is derived from the observed transitions for the radial region from 4 to 11 \AA{} (9 cm$^{-1}$ below the asymptote) and is extended to the longe range region by the use of theoretical dispersion coefficients together with already available photoassociation data. New estimations of the scattering lengths for the complete set of isotopic combinations are derived by mass scaling with the derived potential. The data set for the excited state $2^1\Sigma^+_u$ was sufficient to derive a potential energy curve around the minimum. 
\end{abstract}
\maketitle 

\section{Introduction}

Over the past decades the Sr$_2$ molecule has been the subject of many different studies. In 1977 Miller \cite{Miller1977,Miller1980} discovered by the analysis of matrix isolation spectroscopy two excited $^1\Sigma^+_u$ states belonging to the (5$^1$S + 4$^1$D) and the (5$^1$S + 5$^1$P) asymptotes and investigated the ground state X$^1\Sigma^+_g$. A few years later, in 1980 and 1984, Bergeman \cite{Bergemann1980} and Gerber \cite{Gerber_Sr2_1984} performed heat pipe experiments and studied the transition from the ground state to the 2$^1\Sigma^+_u$ state (5$^1$S + 5$^1$P asymptote), which they called A state. In 1992 Bordas \cite{Bordas1992} discovered by depletion spectroscopy on a molecular beam a higher lying $^1\Pi_u$ state, which was assigned in 1996 by ab initio calculations \cite{Boutassetta} as the 3$^1\Pi_u$ state correlating to the asymptote 5$^3$P+5$^3$P. Several other ab initio calculations were done on Sr$_2$ \cite{Jones1979,OrtizBallone,Wang2000,Czuchaj,Kotochigova2008}. In the most recent calculation \cite{Kotochigova2008} the basis set has been chosen to reproduce best the ground state potential reported by Gerber \cite{Gerber_Sr2_1984}. Since for Strontium the (5$^1$S + 4$^1$D) asymptote lies below the (5$^1$S + 5$^1$P) and thus the A$^1\Sigma^+_u$ state is not the lowest $^1\Sigma^+_u$ state, we renamed this state to 2$^1\Sigma^+_u$ to avoid any further confusion in this work, according to the ab initio calculation \cite{Boutassetta}.

Currently there is high interest on ultracold ensembles of Strontium atoms and high precision spectroscopy on Strontium, because it could be a candidate for an optical frequency standard \cite{Ferrari2003,Santra2005,Boyd2007,Ludlow2008}. Very recently, Zelevinsky \cite{Zelevinsky2008} proposed to measure precisely the time variation of the electron-proton mass ratio by the use of ultracold Sr$_2$ molecules trapped inside an optical lattice. A lot of trap experiments \cite{Mukaiyama,Poli2005,Ferrari2006,FerrariBO2006} at cold and ultracold temperatures are reported for which the reliable knowledge of the interaction properties like scattering lengths of the different isotopes in different electronic states would be of advantage. First estimates of the Strontium scattering lengths for homonuclear collisions of the isotopes $^{88}$Sr and $^{86}$Sr in their electronic ground states are already derived from photoassociation data \cite{Mickelson}. But there is currently no spectroscopic work available, which is sufficiently precise to derive such quantities from the scattering wave-function by directly solving the radial Schroedinger-equation with a complete interaction potential. This would also yield the precise scattering lengths of different isotopes and isotopic compositions, if the Born-Oppenheimer approximation holds for this molecule. 

Our current goal is to derive such precise potentials including the long-range part. We started our work by using the same excitation path as in the earlier measurements \cite{Gerber_Sr2_1984} on the $2^1\Sigma^+_u$ --- X$^1\Sigma^+_g$  transition, where we recorded 37 very dense fluorescence spectra with many overlapping bands. For their analysis we applied a newly written automated software to accelerate the identification and to reduce the probability of transfer error in producing data files for the later fitting routines. 

The paper is divided into the following sections: Section \ref{sec:Experimental} describes the experimental methods and the apparatus, section \ref{sec:Potential} gives the models for data reduction, section \ref{analysis} describes how the analysis and the assignment was done and how the automated software works, section \ref{sec:Results} presents the resulting potentials and Dunham coefficient sets and describes how they were produced and finally section \ref{sec:conclusion} discusses the usability of the presented data including a short outlook.

\section{Experiment}\label{sec:Experimental}

The applied experimental methods are similar to the ones in \cite{Allard_Ca2_2002} for Ca$_2$. To form a gas sample of Sr$_2$ molecules, about 10 g of Strontium are heated inside a stainless steel heat pipe to a temperature of 950 \textcelsius{}. As buffer gas 20 mbar of Argon is introduced. The windows at the ends of the heat pipe are on both sides broad band antireflection coated. 

A special difficulty of the work with alkaline earth metals like Strontium is the relatively high atomic vapor pressure of a few mbar at the melting point of 777 \textcelsius{}, which prevents good heat pipe operation conditions. Thus a significant part of the Strontium vapor condenses in regions of the heat pipe where the temperature is below the melting point and crystals grow towards the center and so towards the path of the laser beam. The time these crystals need to reach the laser beam path strongly depends on the temperature and the buffer gas pressure. Higher temperatures, which would lead to higher production rates of molecules and thereby stronger signals, would also severely shorten the possible measurement time. This can be partially compensated by higher buffer gas pressures, but an increase of the pressure does not only rise the magnitude of pressure broadening and shift, but also leads to additional new problems. Every time Strontium is melting somewhere in the heat pipe or the buffer gas pressure is changed, some kind of fog occurs near the ends of the heat pipe, which mainly consists of fine drops of liquid Strontium, as it became clear when the fog accidentally reached a window and instantly turned it into a metallic layer. This fog produces strong laser stray light, increasing the noise level in the spectra, until it disappears by waiting. The waiting time and the fog density strongly increases with the buffer gas pressure. 

At a temperature of 950 \textcelsius{} and 20 mbar buffer gas it is possible to measure for two to three hours until the Strontium crystals block the laser beam path. When that occurred the heat pipe was shifted inside the oven to heat the ends and melt away the Strontium crystals. By this procedure most of the liquid Strontium gets adsorbed to a mesh and is transported back to the center of the heat pipe, but a smaller part is also transported further to the ends and reaches the windows to a small percentage. Though we did not achieve proper conditions of heat pipe operation where permanent recycling of the material would take place, it was nevertheless possible to record all spectra analyzed for this work without refilling and cleaning the heat pipe. 

The laser excitation is done by a Coherent CR 699 ring dye laser operated with rhodamine 6G in the frequency range from 17000 to 17600 cm$^{-1}$. The fluorescence light emitted antiparallel to the direction of the laser beam is imaged into a Fourier-transform spectrometer (Bruker IFS 120 HR), which is used with a resolution of 0.05 cm$^{-1}$. Most of the recorded spectra are averaged over 10 scans, in few cases also up to 100 scans are taken.  

\begin{figure*}
\resizebox{\textwidth}{!}{%
\includegraphics{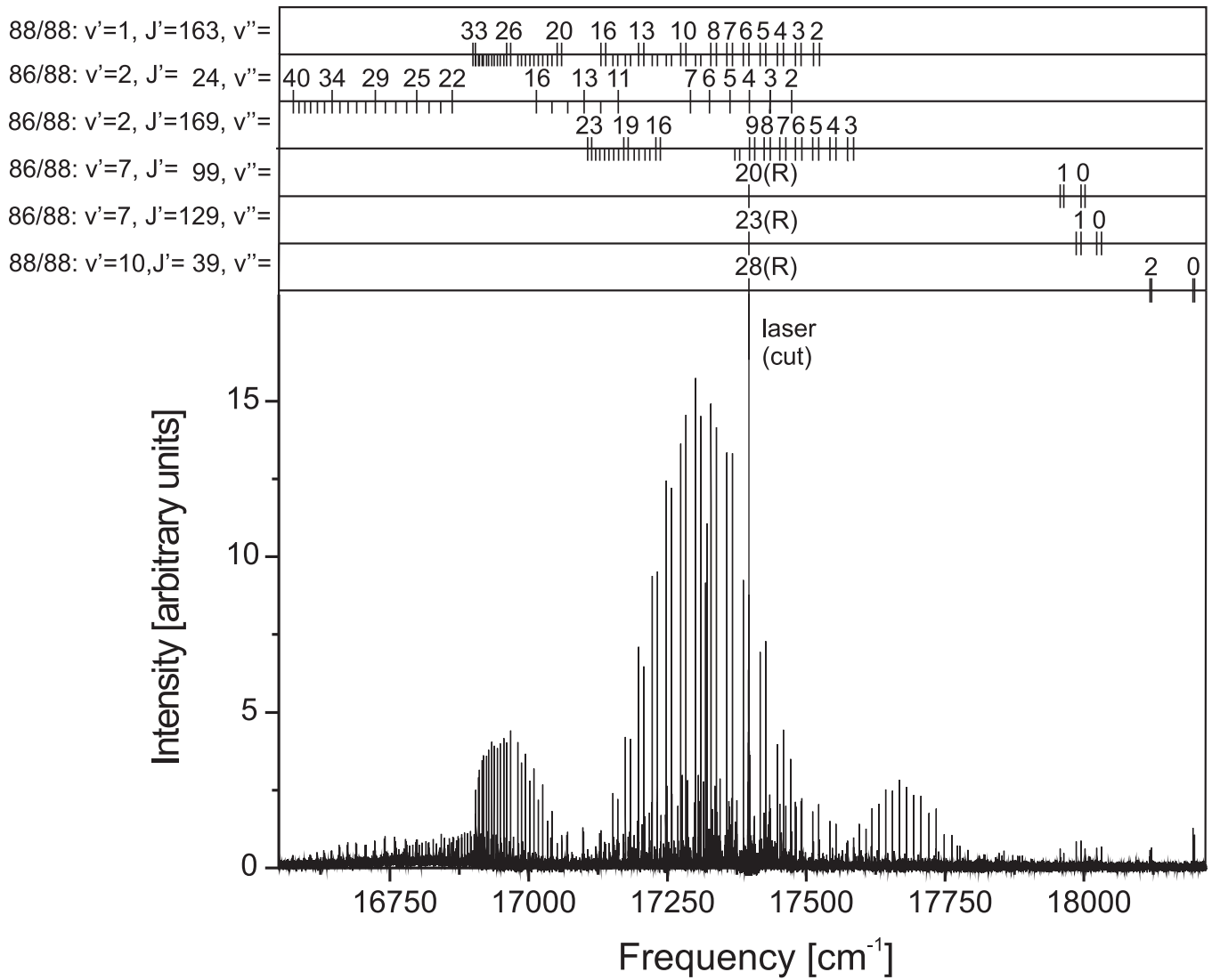}}
\caption{An example spectrum averaged over 50 scans excited by a frequency of $\nu_L=17397.62$ cm$^{-1}$. In total, 14 progressions could be assigned which belong to the $2^1\Sigma^+_u$ --- X$^1\Sigma^+_g$ transition, the quantum numbers of only 6 of them are given on top of the figure including the isotope assignment.}
\label{Spectrum}
\end{figure*}

Figure \ref{Spectrum} shows a typical spectrum. As one can see the density of lines and the amount of excited progressions under single frequency laser excitation (width 20 MHz) is relatively high, resulting from the large reduced mass of the Strontium dimer and the high temperature, which leads to a significant thermal population of all existing rovibrational levels of the flat ground state potential of an excimer type molecule like Sr$_2$.

\section{Potential energy curves and Dunham coefficients}\label{sec:Potential}

The rovibrational energies of electronic states of type $^1\Sigma$ are calculated by the simplest one-dimensional Schr\"odinger equation, which includes the radial kinetic energy, the potential energy curve (PEC) $V(R)$, and the centrifugal energy for the rotational state $J$. $V(R)$ is often called the Born-Oppenheimer potential because no corrections for the coupling between the nuclear and the electronic motion is introduced in this simplest form of the molecular Schr\"odinger equation. For the PEC $V(R)$ the same analytical representation is used as e.g. in our earlier work on Ca$_2$ \cite{Allard_Ca2_2002}: 

It is divided into three parts: The central part of the potential ($R_i\leq R\leq R_a$) is given by

\begin{equation}
\label{GVC}
V_c(R)=T_m+\sum_ia_ix^i
\end{equation} 

\noindent with 

\begin{equation}
\label{GVx}
x=\frac{R-R_m}{R+bR_m} \mbox{ .}
\end{equation}

The inner repulsive wall ($R\leq R_i$) is represented by

\begin{equation}
\label{GVi}
V_i(R)=A+\frac{B}{R^n} \mbox{ .}
\end{equation}

The long range part ($R\geq R_a$) is described by

\begin{equation}
\label{GVa}
V_a(R)=U_\infty-\frac{C_3}{R^3}-\frac{C_6}{R^6}-\frac{C_8}{R^8}-\frac{C_{10}}{R^{10}} \mbox{ .}
\end{equation}

Here the $a_i$ are the fitting parameters. The radius $R_m$ forms together with $T_m$ the expansion point of the potential and is typically chosen to be close to the equilibrium distance. Here it is taken as the radius of the minimum of a prior fitting result. The connection radii $R_i$ and $R_a$, as well as the parameters b and n are manually adjusted to get an acceptably small standard deviation using a low number of fit parameters $a_i$. The application of the long range parameters slightly differs for the ground state and the excited state (see sec. \ref{sec:Results} for details), the parameter C$_3$ is needed only for the excited state, while the parameters C$_6$ to C$_{10}$ are used for the ground state. 

Since the fit of this kind of potential description has a slow convergence, for the assignment of the data and to obtain RKR potentials as good starting potentials, a set of Dunham coefficients is derived first \cite{Townes}:

\begin{equation}
\label{GDun}
E^i_{vJ}=T+\sum_{k,l}Y_{lk}\left(\sqrt{\frac{\mu_0}{\mu_i}}\right)^{l+2k}{\left(v+\frac{1}{2}\right)}^l\cdot{\left[J(J+1)\right]}^k
\end{equation}

The $Y_{lk}$ are the Dunham parameters for the chosen reference isotopomer $^{88}$Sr$_2$ with reduced mass $\mu_0$. T is the origin of the state. The reference of energy levels is defined by setting T for the ground state X$^1\Sigma^+_g$ to zero. The $\mu_i$ is the reduced mass of the isotopomer $i$, for which the energy ladder is considered. 

\section{Analysis of spectra}
\label{analysis}

The main part of the analysis of data is done using a partially automated software, of which a prior version is already mentioned in \cite{SteinLiCs2008}. First attempts of assigning the spectra were done using the set of Dunham coefficients reported in \cite{Gerber_Sr2_1984} together with a simple program, where the user has to select at least three lines from a possible progression, for which the program subsequently suggests assignments and automatically searches for additional lines within the spectrum. Using this function it became clear within short time, that the assumed rotational dependence of the term energies calculated with the Dunham set from \cite{Gerber_Sr2_1984} is incorrect even considering the relatively low accuracy of a few tenths of a wavenumber as given in \cite{Gerber_Sr2_1984}. In many cases the vibrational energy differences of odd rotational ground state levels of the reference isotopomer $^{88}$Sr$_2$ were suggested as the ones fitting best, though these levels do not exist for homonuclear isotopomers with nuclear spin zero like the even isotopes of Sr. In the diploma thesis by Schneider \cite{SchneiderDip}, from which the paper \cite{Gerber_Sr2_1984} was derived, one can read that only the three strongest doublet progressions were used for the rovibrational assignment of the ground state and for the fitting of the Dunham-coefficient set which they used for the calculation of the RKR potential. Obviously their selected data set was too small to fix the assignment unambiguously, so our assignment procedure was started from the beginning.

\subsection{The initial assignment}
\label{initAss}

Initially, three progressions belonging to the same upper v$'$ and neighboring J$'$ quantum numbers are taken, which were specifically produced by exciting at wavelengths where collisionally induced rotational satellite lines appear. In this way the relative J and v assignment was fixed and only an offset in J of the whole set had to be varied to find the correct absolute rotational assignment. This was first done in a combined fit of Dunham coefficients for both electronic states, which was possible, because the rovibrational levels of the upper 2$^1\Sigma^+_u$ state do not show local perturbations in the small energy interval studied here. For the standard deviation $\sigma$, which was minimized, we use the conventional definition:

\begin{equation}
\label{sigma}
\sigma^2=\frac{1}{N-M}\sum_{n=1}^N\frac{\left(E_{obs,n}-E_{calc,n}\right)^2}{\rho_n^2}
\end{equation}

Here N is the number of observations, M the number of fit parameters, and $\rho_n$ the uncertainty for the nth term energy or transition frequency.

Among the first three measured progressions there was one which was already shown in \cite{Gerber_Sr2_1984} with the assignment (v$'=1$, J$'=37$). Varying the J offset of these three progressions yielded with almost the same quality three assignments, namely J$'=37$, 39, or 41. After adding a fourth such progression to the data set, only an assignment to J$'$=39 or 41 remained possible, and after adding a fifth only for J$'=41$ the standard deviation $\sigma$ was below 1.0 for an estimated uncertainty of $\rho=0.005$ cm$^{-1}$ for each line. This rotational assignment then became unambiguous after adding more and more excitations to the data set. The data set was extended to higher J values and to different upper vibrational levels by stepwise adding progressions.

The vibrational assignment was taken from \cite{Gerber_Sr2_1984} and could later be confirmed by the observations of progressions belonging to the isotopomers $^{86}$Sr$^{88}$Sr and $^{87}$Sr$^{88}$Sr which were consistently described with the Dunham set by mass scaling (see eq. \ref{GDun}). Also the results of the potential fits, which are described in Sec. \ref{sec:Results}, confirm this vibrational assignment. The v$'$ assignment of the excited state is also confirmed by the intensity envelopes of the progressions e.g. originating from the lowest vibrational level of the 2$^1\Sigma^+_u$ state, i.e. v$'=0$, the envelope shows a single maximum. Going to higher v$'$ one sees as simple image of the wave function an envelope with the appropriate number of maxima, v$'+1$.

\subsection{The automated assignment of the main body of data} 

Besides the description of the ground state also an analysis of the excited state is developed. But each laser excited progression only gives information about one single rovibrational level of the excited state, so the information contained in the huge amount of weak lines belonging to overlapping excited progressions is very valuable, too. Since the amount of collisionally induced rotational satellite lines, which are intensively used, e.g. in \cite{SteinLiCs2008,Docenko_NaRb2005,Docenko_NaCs3Pi}, for the description of the excited states, is very low in this case, caused by the very weak overall intensity of the Sr$_2$ fluorescence spectra, the analysis of these overlapping excitations is of special importance. Doing such an analysis by hand or even applying the half automated method as for the initial assignment of the strongest lines would have been very error prone and extremely time consuming, since it is not straightforward to decide which lines belong to the same progression. Even for a lot of combinations of lines, which turn out later not to belong to the same progression, there is often more than one choice where they would fit nearly equally well into the ground state term energy set.

To do the assignment with high reliability and with many consistency checks within limited time a special program code was written. The algorithm basically checks for every unassigned line, starting with the strongest one, every possibility of assignment. First it is checked if the possible upper level could be excited by the actual laser setting or alternatively if it could be a collisionally induced satellite of an already assigned stronger progression. This information is collected for the present recording and the correctness of the assumed assignment is mainly judged by the number of lines found inside the spectrum which possibly belong to a common progression, and in the case of a P/R doublet progression by the number of complete doublets. An important criterion is the maximal length of a progression, i.e. there is no assignment found which gives more Q lines or more complete P/R doublets in the recording. An additional acceptance criterion is the positive answer to the question if the upper level of the selected progression also fits into the simultaneously generated description of the excited state 2$^1\Sigma^+_u$.

In the final assignment run of an iterative series of assignments, of Dunham fits, and of potential fits (see section \ref{sec:Potential}), it was possible to identify around 10300 lines from the 37 recorded spectra within a computation time of about 8 hours. The resulting data confirm the applied model, because it describes them by the potentials and sets of Dunham coefficients with standard deviations $\sigma$ (see eq. \ref{sigma}) smaller than one. The uncertainties $\rho$ are mostly four times smaller than the deviation limit set for the predicted value during the assignment, 0.02 cm$^{-1}$ for the ground state and 0.04 cm$^{-1}$ for the excited state. Also very weak progressions of lines with signal-to-noise ratios down to 2.5 could be reliably assigned due to the many consistency checks. Only few single strong lines and some weak lines with S/N below 7 and very few medium strong lines with S/N below 30 remain unassigned. The few unassigned strong lines form a doublet together with the laser line and are clearly identifiable as molecular lines because of rotational satellite structures, but do not fit into the Sr$_2$ ground state potential as a P/R doublet. These lines possibly belong to another molecule as for example SrH or SrO. Few remaining unassigned lines could belong to different excited states, e.g. the lower 1$^1\Sigma^+_u$ state or the lowest two $^1\Pi_u$ states obtained by theoretical calculations \cite{Boutassetta,Czuchaj,Kotochigova2008}, but which have, however, not been investigated so far. The main part of unassigned lines probably belongs to progressions which are too weak to match the requested assignment criterion of at least two clearly visible P/R doublets for each progression. 

\subsection{Resulting data fields}\label{sec:datafields}

\begin{figure}
\resizebox{0.5\textwidth}{!}{%
\includegraphics{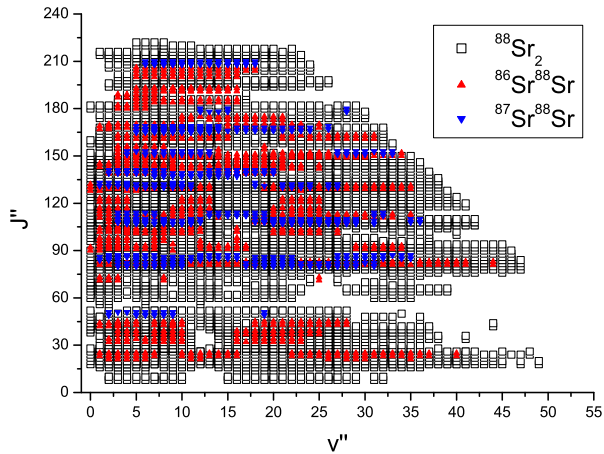}}
\caption{Overview of the observed energy levels of the ground state X$^1\Sigma^+_g$.}
\label{XData}
\end{figure}

\begin{figure}
\resizebox{0.5\textwidth}{!}{%
\includegraphics{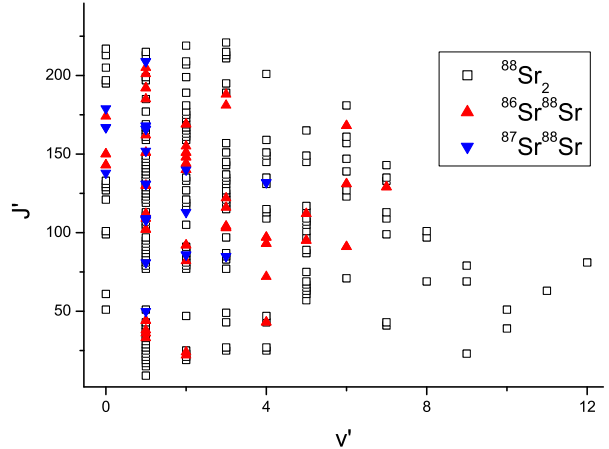}}
\caption{Observed energy levels of the excited state $2^1\Sigma^+_u$.}
\label{CData}
\end{figure}

The data set obtained consists of about 4673 rovibrational energy levels for the ground state and 260 for the excited state 2$^1\Sigma^+_u$. For the ground state 3163 levels belong to the isotopomer $^{88}$Sr$_2$, 1059 to the isotopomer $^{86}$Sr$^{88}$Sr and 451 to $^{87}$Sr$^{88}$Sr. The data fields are shown in figure \ref{XData} for the ground state and figure \ref{CData} for the excited state. 

Within a single progression the relative uncertainties of the transitions are estimated to be mainly about $0.005$ cm$^{-1}$ from the applied resolution (0.05 cm$^{-1}$) and the obtained S/N. The Doppler width is 0.03 cm$^{-1}$ for a temperature of 950 \textcelsius{} and a frequency of 17250 cm$^{-1}$. Since overlapping excitations are evaluated, which contribute the by far largest amount of transitions to the data set, the excitation is not always on resonance and so lines of molecules with non zero velocity components parallel to the direction of observation are very likely. Thus Doppler shifts cannot be excluded. However, for the ground state only the differences of possible Doppler shifts of the individual lines within a progression contribute to the uncertainty. The depth of the ground state potential is below 1100 cm$^{-1}$, so the shift here is well below 0.005 cm$^{-1}$. The relative accuracy of the Fourier-transform spectrometer is 0.001 cm$^{-1}$. The errors of individual lines, which overlap with other transitions, are enlarged to 0.01 cm$^{-1}$. 

The term energies of the excited 2$^1\Sigma^+_u$ state are derived from the term energies of the ground state calculated from the X state potential (see section \ref{sec:Results}) by adding the measured transition frequencies, and averaged over all observations of the same level. The uncertainties for the resulting energies are estimated to be 0.01 cm$^{-1}$. Since here the absolute transition frequencies are needed for determining the excited state levels the Doppler effect is the main source of error. Because of the overall weakness of the observed lines the probability of excitation in the wings of the Doppler-profile is rather low, so the estimation of an uncertainty of 0.01 cm$^{-1}$ should be a conservative limit.

The obtained standard deviations $\sigma$ (eq. \ref{sigma}) of the potential and the Dunham fits show that the given errors are rather overestimated than underestimated.


\section{Results}\label{sec:Results}

\subsection{The X$^1\Sigma^+_g$ ground state}

\begin{table}
\caption{Dunham coefficients for the X$^1\Sigma^+_g$ ground state. All values are in cm$^{-1}$ and for the reference isotopomer $^{88}$Sr$_2$. For the range of quantum numbers see text.}
\label{X1Sigma+gDunham}
\centering
\begin{tabular}{rrrr}
\hline\noalign{\smallskip}
\(l\downarrow{} k\rightarrow{}\) & 0 & 1 & 2\\ \hline\noalign{\smallskip}
0 &    0   & 0.0175795 & -1.4016$\times 10^{-8}$\\
1 & 40.32831 & -0.000168 & -2.1935$\times 10^{-10}$\\
2 & -0.39943 & -1.037$\times 10^{-6}$ & -2.052$\times 10^{-11}$\\
3 &     0  & 4.55$\times 10^{-8}$ &   0   \\
4 & 1.609$\times 10^{-5}$ & -5.521$\times 10^{-9}$ & 6.0206$\times 10^{-14}$\\
5 & -4.802$\times 10^{-7}$ & 2.8508$\times 10^{-10}$ &    0  \\
6 & 1.3925$\times 10^{-8}$ & -7.5518$\times 10^{-12}$ & -1.7023$\times 10^{-16}$\\
7 & -2.49438$\times 10^{-10}$ & 9.7704$\times 10^{-14}$ & 5.1207$\times 10^{-18}$\\
8 & 1.86553$\times 10^{-12}$ & -4.992$\times 10^{-16}$ & -4.517$\times 10^{-20}$\\
\hline
\hline\noalign{\smallskip}
\(l\downarrow{} k\rightarrow{}\) & 3 & 4 & 5\\ \hline\noalign{\smallskip}
0 &    0   &   0    & -4.416$\times 10^{-24}$\\
1 & -4.406$\times 10^{-15}$ &   0    & 6.2378$\times 10^{-25}$\\
2 & 9.032$\times 10^{-16}$ & -1.5006$\times 10^{-20}$ &  0    \\
3 & -5.443$\times 10^{-17}$ & 9.8001$\times 10^{-22}$ &   0   \\
4 &    0   &  0     & -3.1324$\times 10^{-28}$\\
5 & 0      &    0   &      \\
6 &    0   & 1.2277$\times 10^{-26}$ &      \\
7 & 5.5692$\times 10^{-23}$ & -7.575$\times 10^{-28}$ &      \\
8 & -1.0852$\times 10^{-24}$ &       &      \\
\hline
\end{tabular}
\end{table}

The final set of Dunham coefficients with 35 freely variable parameters for the ground state is constructed by fitting the calculated level differences to differences of transition frequencies within the individual progressions. The result can be found in table \ref{X1Sigma+gDunham} and describes those rovibrational levels shown in fig. \ref{XData}, which are on the left side of a line defined by the points (v$''$=47, J$''$=79) and (v$''$=17, J$''$=219). The standard deviation is $\sigma=0.77$. For the description of the asymptotic energy levels by Dunham coefficients an inappropriately large amount of coefficients would be necessary and even a convergence is not guaranteed. The Dunham parameters should be used with caution, a physical meaning can only be attributed to the lowest ones like $Y_{10}$, $Y_{01}$ and $Y_{02}$.

For the fit of the X state potential curve (see section \ref{sec:Potential} for the description of the analytical model) the energies of the originating excited levels of the progressions are handled as free parameters. In this fit the energies are calculated with reference to the ground state asymptote. Therefore, the potential parameter $U_\infty$ is set to 0 and the parameter $T_m$ is adjusted to get a continuous connection of the central potential part ($R_i\leq R\leq R_a$) to the long range area ($R\geq R_a$). The parameter $C_3$ is 0 for the ground state while the parameters $C_6$ and $C_8$ are taken from theory \cite{Porsev2006}. To improve the description of the data the parameter $C_{10}$ (also theoretical values exist) was varied. The parameter $A$ is adjusted to get a continuous connection to the inner potential wall ($R\leq R_i$, formula (\ref{GVi})), while the parameter $B$ is varied in the potential fit. 

\begin{figure}
\resizebox{0.5\textwidth}{!}{%
\includegraphics{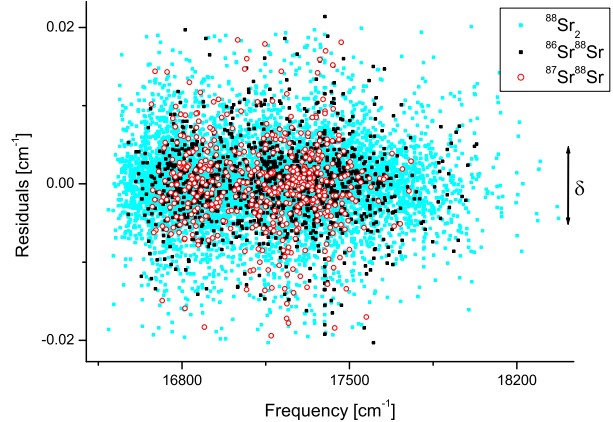}}
\caption{The residuals of the X$^1\Sigma^+_g$ potential fit as a function of the transition frequencies. The range $\delta$ marks the uncertainty interval estimated for most lines.}
\label{XResiduals}
\end{figure}

\begin{table}
\caption{Potential coefficients for the X$^1\Sigma^+_g$ ground state, referenced to the asymptote.}
\label{X1Sigma+gPotential}
\centering
\centering
\begin{tabular}{lp{0.1\textwidth}r}\hline
a$_{1}$   && -1.47$\times 10^{-2}$ cm$^{-1}$ \\
a$_{2}$   && 4.2868277$\times 10^{3}$ cm$^{-1}$ \\
a$_{3}$   && -1.26953$\times 10^{2}$ cm$^{-1}$ \\
a$_{4}$   && -3.8754510$\times 10^{3}$ cm$^{-1}$ \\
a$_{5}$   && 3.410482$\times 10^{3}$ cm$^{-1}$ \\
a$_{6}$   && 7.9906800$\times 10^{3}$ cm$^{-1}$ \\
a$_{7}$   && -1.52520496$\times 10^{5}$ cm$^{-1}$ \\
a$_{8}$   && 4.056374$\times 10^{3}$ cm$^{-1}$ \\
a$_{9}$   && 1.975774038$\times 10^{6}$ cm$^{-1}$ \\
a$_{10}$   && -1.722492339$\times 10^{6}$ cm$^{-1}$ \\
a$_{11}$   && -1.366495056$\times 10^{7}$ cm$^{-1}$ \\
a$_{12}$   && 2.180878197$\times 10^{7}$ cm$^{-1}$ \\
a$_{13}$   && 4.6759142229$\times 10^{7}$ cm$^{-1}$ \\
a$_{14}$   && -1.1928451623$\times 10^{8}$ cm$^{-1}$ \\
a$_{15}$   && -4.105311327$\times 10^{7}$ cm$^{-1}$ \\
a$_{16}$   && 3.0056913041$\times 10^{8}$ cm$^{-1}$ \\
a$_{17}$   && -1.6227920475$\times 10^{8}$ cm$^{-1}$ \\
a$_{18}$   && -2.4953885924$\times 10^{8}$ cm$^{-1}$ \\
a$_{19}$   && 3.2507145429$\times 10^{8}$ cm$^{-1}$ \\
a$_{20}$   && -1.086437025$\times 10^{8}$ cm$^{-1}$ \\
\hline
b     && -0.57    \\
R$_m$ && 4.67169686 \AA \\ 
T$_m$ && -1081.8163 cm$^{-1}$ \\ \hline
R$_i$ && 3.98 \AA \\
n     && 6    \\
A     && -2.5412398$\times 10^{3}$ cm$^{-1}$ \\
B     && 9.88561567$\times 10^{6}$ cm$^{-1}$\AA$^6$ \\ \hline
R$_a$ && 11.0 \AA \\
C$_{6}$ \cite{Porsev2006} && 1.4955$\times 10^{7}$ cm$^{-1}$\AA$^{6}$ \\
C$_{8}$ \cite{Porsev2006} && 5.1175$\times 10^{8}$ cm$^{-1}$\AA$^{8}$ \\
C$_{10}$ && 2.495$\times 10^{10}$ cm$^{-1}$\AA$^{10}$ \\
U$_\infty$ && 0.0 cm$^{-1}$ \\ \hline
\multicolumn{2}{l}{derived constants:}       \\
D$_e$    && 1081.82 cm$^{-1}$ \\
R$_e$    && 4.67174 \AA \\ \hline
\end{tabular}
\end{table}

Figure \ref{XResiduals} shows the residuals of the potential fit as a function of the energy, while the resulting potential coefficients (only 22 free parameters compared to 35 Dunham coefficients) are given in table \ref{X1Sigma+gPotential}. The weighted standard deviation of the fit is $\sigma$ = 0.81. The potential was applied for the integration of the Schr\"odinger equation in the interval from 3.3 \AA{} to 100 \AA{}.

To improve the description of the long range part we also used information from trap experiments \cite{Mickelson,Yasuda} derived by photoassociation. In these experiments the photoassociation laser was tuned to excite high vibrational levels of the $2^1\Sigma^+_u$ state. Such resonances lead to trap losses the magnitude of which is directly related to the Franck-Condon density (the square of the overlap integral of the scattering wave function of the ground state and the vibrational wave function of the $2^1\Sigma^+_u$ state) of the excited transitions. Because the vibrational wave functions of bound states have significant amplitudes mainly close to the outer turning points, the intensity envelope of the photoassociation resonances gives information about the amplitude of the scattering wave function and thus about the node positions in this region. Here the position of the second to last node for the molecule $^{86}$Sr$_2$ was obtained by \cite{Mickelson} to be at 62.6 a$_0$ and the position of the last node for the molecule $^{88}$Sr$_2$ slightly below 40 \AA{} by \cite{Yasuda}. 

To include this information in our potential determination, we first adjusted the C$_{10}$ coefficient to fit the position of the second to last node of the isotopomer $^{86}$Sr$_2$ to the value above and calculated the term energy for the last vibrational level (J$''$=0, v$''$=62) with this potential. This term value was included in a new overall fit as an observed quantity with an error of $3\times 10^{-5}$ cm$^{-1}$. Finally, it was checked that this potential also reproduced quite well the position of the last node of the scattering wave function of $^{88}$Sr$_2$ reported by Yasuda \cite{Yasuda}. 

\begin{figure}
\resizebox{0.5\textwidth}{!}{%
\includegraphics{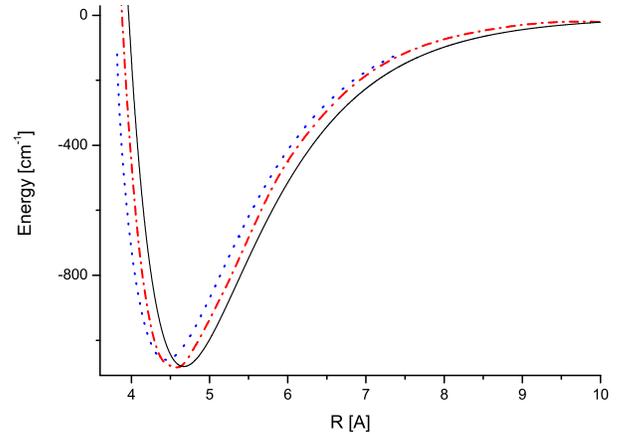}}
\caption{The X$^1\Sigma^+_g$ ground state potential (solid line) compared to the RKR-potential from \cite{Gerber_Sr2_1984} (dotted) and the ab initio potential from \cite{Boutassetta} (dashed dotted).}
\label{XPotential}
\end{figure}

Figure \ref{XPotential} shows the final potential, which is valid in the region from 4 \AA{} to 11 \AA{} for our spectroscopic data, compared to the RKR potential \cite{Gerber_Sr2_1984} and the theoretical calculations \cite{Boutassetta}. Here it is clearly visible, that the deviations of the two experimental potentials caused by the change in the rotational assignment are mostly bigger than the differences to the theoretical calculations. 


\subsection{The excited 2$^1\Sigma^+_u$ state}

\begin{table}
\caption{Dunham coefficients for the excited state 2$^1\Sigma^+_u$, for the isotopomer $^{88}$Sr$_2$. All values in cm$^{-1}$. T is referenced to the origin of the Dunham series for the ground state.}
\label{C1Sigma+uDunham}
\centering
\begin{tabular}{rrrr}
\hline\noalign{\smallskip}
\(l\downarrow{} k\rightarrow{}\) & 0 & 1 & 2\\ \hline\noalign{\smallskip}
0 &   0    & 0.0219691 & -5.962$\times 10^{-9}$\\
1 & 84.207 & -6.693$\times 10^{-5}$ & -6.17$\times 10^{-11}$\\
2 & -0.2639 & -6.38$\times 10^{-7}$ &      \\
3 & -0.001422 &       &      \\
\hline
\hline\noalign{\smallskip}
$T$ & 17358.7262 &     &   \\
\hline
\end{tabular}
\end{table}

Table \ref{C1Sigma+uDunham} shows the set of Dunham coefficients with 9 freely varied parameters derived in a combined fit for both electronic states from 260 levels of the low part of the 2$^1\Sigma^+_u$ state. It reproduces all observed energy levels with v$'$ up to 12 and J from 9 to 221 with a standard deviation of $\sigma=0.80$.

For the final potential description of the excited state 2$^1\Sigma^+_u$ term energies are derived, which are referenced to the potential minimum of the ground state. As already mentioned in section \ref{sec:datafields}, these term energies are calculated by adding the measured transition frequencies to the term energies calculated from the ground state potential. The model description differs slightly from that of the ground state. Here the parameter $T_m$ is included in the fit while the parameter $U_\infty$ is calculated by adding the ground state dissociation energy of 1081.82 cm$^{-1}$ to the atomic transition frequency for $^1$S$_0 \longrightarrow ^1$P$_1$ of 21698.482 cm$^{-1}$ obtained from \cite{Moore}. The long range coefficient $C_3$ is taken from \cite{Yasuda}, while the $C_{6}$ and $C_{8}$ coefficients are adjusted to get a continuously differentiable connection of the long range part ($R\geq R_a$) to the central region ($R_i\leq R\leq R_a$). The parameters $A$ and $B$ are adjusted to get a continuously differentiable connection of the inner potential wall ($R\leq R_i$). 

The 2$^1\Sigma^+_u$ potential supports roughly 350 bound vibrational levels, with a large uncertainty, $\pm$ 50 or more, because the gap between our data and the photoassociation data \cite{Yasuda,Yasuda_private} is about 4000 cm$^{-1}$.

\begin{table}
\caption{Potential coefficients for the excited state 2$^1\Sigma^+_u$ with reference to the minimum of the X state.}
\label{C1Sigma+uPotential}
\centering
\begin{tabular}{lp{0.1\textwidth}r}\hline
a$_{1}$   && 1.619$\times 10^{0}$ cm$^{-1}$ \\
a$_{2}$   && 2.0176667$\times 10^{4}$ cm$^{-1}$ \\
a$_{3}$   && 1.049518$\times 10^{4}$ cm$^{-1}$ \\
a$_{4}$   && -1.47839$\times 10^{4}$ cm$^{-1}$ \\
a$_{5}$   && -4.65051$\times 10^{4}$ cm$^{-1}$ \\
a$_{6}$   && -5.36901$\times 10^{4}$ cm$^{-1}$ \\
\hline
b     && -0.50    \\
R$_m$ && 4.1783479 \AA \\ 
T$_m$ && 17358.7389 cm$^{-1}$ \\ \hline
R$_i$ && 3.775 \AA \\
n     && 6    \\
A     && 1.4855497$\times 10^{4}$ cm$^{-1}$ \\
B     && 1.01060899$\times 10^{7}$ cm$^{-1}$\AA$^6$ \\ \hline
R$_a$ && 5.0 \AA \\
C$_{3}$ \cite{Yasuda} && 5.9712$\times 10^{5}$ cm$^{-1}$\AA$^{3}$ \\
C$_{6}$ \footnotemark[1] && -5.1541$\times 10^7$cm$^{-1}$\AA$^6$ \\
C$_{8}$ \footnotemark[1] && 8.9977$\times 10^{8}$cm$^{-1}$\AA$^8$ \\
U$_\infty$ && 22780.30 cm$^{-1}$ \\ \hline
\multicolumn{2}{l}{derived constants:}       \\
T$_e$    && 17358.739 cm$^{-1}$ \\
R$_e$    && 4.17828 \AA \\ \hline
\end{tabular}
\footnotetext[1]{The coefficients C$_6$ and C$_8$ are only parameters for connection to the atomic asymptote $^1$S$_0+^1$P$_1$.}
\end{table}

\begin{figure}
\resizebox{0.5\textwidth}{!}{%
\includegraphics{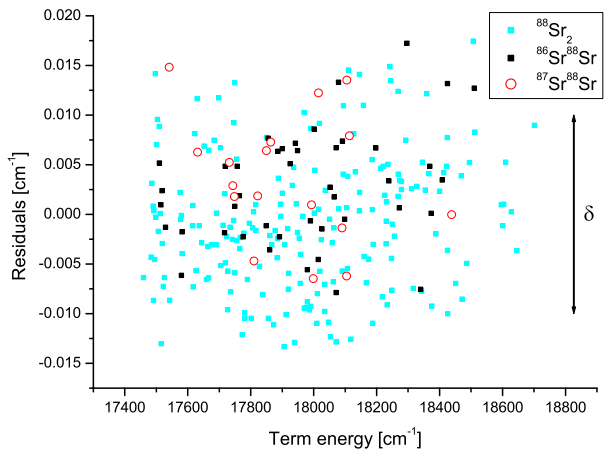}}
\caption{The residuals of the 2$^1\Sigma^+_u$ potential fit as a function of the term energies.}
\label{CResiduals}
\end{figure}

\begin{figure}
\resizebox{0.5\textwidth}{!}{%
\includegraphics{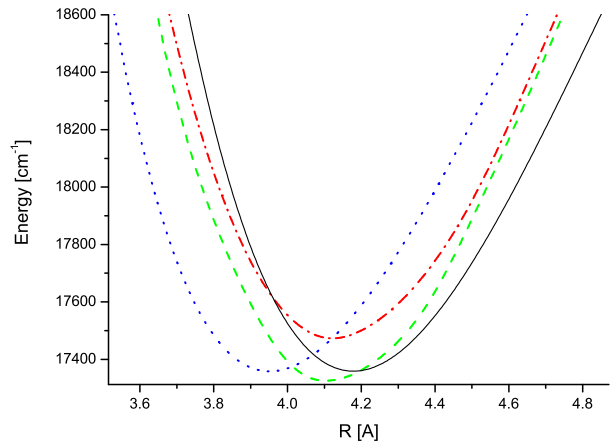}}
\caption{The derived potential (solid line) compared to the RKR potential from Gerber \cite{Gerber_Sr2_1984} (dotted) for the state 2$^1\Sigma^+_u$ and the ab initio potentials from Boutassetta \cite{Boutassetta,Frecon_priv} (dashed dotted) and Czuchaj \cite{Czuchaj,Krosnicki_priv} (dashed).}
\label{CPotential}
\end{figure}

The residuals of the potential fit, which has a weighted standard deviation of $\sigma$ = 0.68 (defined in eq. (\ref{sigma})), can be found in figure \ref{CResiduals}. 
The potential coefficients are listed in table \ref{C1Sigma+uPotential}. The integration interval used for solving the radial Schr\"odinger equation extends again from 3.3 to 100 \AA.


Figure \ref{CPotential} gives a comparison of the resulting potential energy curve (PEC) to the experimental potential from \cite{Gerber_Sr2_1984} and to the theoretical calculations \cite{Boutassetta} and \cite{Czuchaj}. Here the large deviations between the two experimental potentials are again obvious, due to the different rotational assignments of the excited state, which are directly related to the ground state.

\section{Conclusion and outlook}\label{sec:conclusion}

In this article two electronic states of the Sr$_2$ dimer were studied, the ground state X$^1\Sigma^+_g$ and the minimum region of the excited state 2$^1\Sigma^+_u$. During the analysis the rotational assignment had to be shifted by four units upwards compared to the previously most precise investigation of this molecule \cite{Gerber_Sr2_1984}. This also implies a significant shift of the potentials towards larger inter atomic distances. The observed 3163 ground state levels for the isotopomer $^{88}$Sr$_2$ are more than half of the existing rovibrational bound levels of this isotopomer (around 5400). The observed ground state levels of the isotopomers $^{86}$Sr$^{88}$Sr (1059) and $^{87}$Sr$^{88}$Sr (451) further confirm the new rotational and the earlier vibrational assignment. 

\begin{table}
\caption{The scattering lengths for the different isotopic combinations calculated using the long range coefficients from \cite{Mitroy2003} and from \cite{Porsev2006}, all values in atomic units (a$_0 = 0.5292$ \AA).}
\label{Scat}
\centering
\begin{tabular}{rcc}
\hline\noalign{\smallskip}
isotopomer & fit using \cite{Mitroy2003} & fit using \cite{Porsev2006} \\
$^{84}$Sr+$^{84}$Sr & 121    & 127    \\
$^{84}$Sr+$^{86}$Sr & 30     & 36     \\
$^{84}$Sr+$^{87}$Sr & -64    & -45    \\
$^{84}$Sr+$^{88}$Sr & 1170   & -1900 \footnotemark \\
$^{86}$Sr+$^{86}$Sr & 677    & 1430   \\
$^{86}$Sr+$^{87}$Sr & 160    & 171    \\
$^{86}$Sr+$^{88}$Sr & 97     & 101    \\
$^{87}$Sr+$^{87}$Sr & 95     & 99     \\
$^{87}$Sr+$^{88}$Sr & 54     & 58     \\
$^{88}$Sr+$^{88}$Sr & -4.8   & 4.5    \\
\hline
\end{tabular}
\footnotetext{For the isotopic combination $^{84}$Sr+$^{88}$Sr the number of vibrational levels decreases from 63 to 62 when the long range coefficient set from \cite{Porsev2006} is used instead of that from \cite{Mitroy2003}, thus only a large magnitude is predicted here. Also the sign in the case of $^{88}$Sr+$^{88}$Sr is not determined, while in this case the scattering length is close to zero.}
\end{table}

We improved the asymptotic range of the ground state potential by the use of photoassociation data \cite{Mickelson,Yasuda,Yasuda_private} and calculated the scattering lengths for all combinations of natural Strontium isotopes. However, there remains still a significant uncertainty which stems from the uncertainty in the long range coefficients, especially from that of the C$_6$ coefficient, as discussed in \cite{Mickelson}. Regarding this coefficient five different theoretical calculations \cite{Stanton1994,Porsev2002,Mitroy2003,Lima2005,Porsev2006} were published during the recent years, with values ranging from 3103 a.u. \cite{Porsev2006} to 3249 a.u. \cite{Mitroy2003}. Since just in these two publications with  results deviating strongest the authors have used the most advanced methods and are the only ones who give also C$_8$ and C$_{10}$ coefficients, we performed two different potential fits applying both sets of C$_i$. The first one using the coefficients from \cite{Porsev2006} is shown in table \ref{X1Sigma+gPotential}.

The dissociation energy $D_e$ differs for both fits: Using the coefficients from \cite{Porsev2006} we obtained $D_e = 1081.82 $ cm$^{-1}$, and with those from \cite{Mitroy2003} $D_e = 1081.52$ cm$^{-1}$. In fits to reproduce the scattering node position from \cite{Mickelson} for the potential using the coefficients from \cite{Mitroy2003} the coefficient C$_{10}$ had to be reduced from $4.25\times 10^7$ a.u. to $1.52\times 10^7$ a.u. while for the potential using the coefficients \cite{Porsev2006} it had to be increased from $4.22\times 10^7$ a.u. to $6.60\times 10^7$ a.u. We have no argument which solution should be preferred, thus we need more measurements. The number of vibrational levels, 62 for the three lighter isotopomers  $^{84}$Sr$_2$, $^{86}$Sr$^{84}$Sr and $^{87}$Sr$^{84}$Sr and 63 for the six heavier isotopomers, is the same in both cases, only for $^{84}$Sr$^{88}$Sr it changes from 62 for the long range coefficients from \cite{Mitroy2003} to 63 for the coefficients from \cite{Porsev2006}, respectively.

The resulting scattering lengths are compared in table \ref{Scat}. Here the two results for the isotopic combination $^{86}$Sr+$^{86}$Sr are clearly located inside the range given in \cite{Mickelson} (610 a$_0$ to 2300 a$_0$), as expected because of the use of their node position of the scattering wave function, but for the combination $^{88}$Sr+$^{88}$Sr the value of -4.8 a$_0$ calculated using the long range coefficients from \cite{Mitroy2003} is outside the range of -1 a$_0$ to 13 a$_0$ reported in \cite{Mickelson}. Mickelson calculated \cite{Mickelson} the scattering length for the isotopic combination $^{88}$Sr+$^{88}$Sr by mass scaling from the result for the combination $^{86}$Sr+$^{86}$Sr using either the ab initio potential \cite{Boutassetta} or the RKR potential \cite{Gerber_Sr2_1984}, which both support a lower number of vibrational levels compared to our result. 

Another uncertainty for the prediction of the scattering lengths lies in the accuracy of the Born-Oppenheimer approximation for this molecule, though possible corrections necessary for mass scaling should be small because of the large mass of the atoms.

The derived potentials present an excellent starting point for further spectroscopic investigations of this molecule, concentrating on a more complete investigation of the long range part of the ground state. This will help to reduce the uncertainties in the calculation of the scattering lengths. The calculated Franck-Condon factors for the transition 2$^1\Sigma^+_u$ --- X$^1\Sigma^+_g$ direct us that the highest vibrational levels for $^{88}$Sr$_2$ could be observed through fluorescence by laser excitation of the v$'$=4 level of the state 2$^1\Sigma^+_u$ starting from a low vibrational level of the ground state as e.g. v$''=2$. An investigation is presently prepared. 
If this should not give satisfying results, a different technique will be applied, where the v$'$=4 level becomes excited directly from the desired asymptotic levels, which are significantly thermally populated due to the high temperatures and the flat ground state potential. Because of the high density of vibrational levels close to the ground state asymptote the number of overlapping bands will be very high, too. A monochromator could be used to filter the fluorescence in the range where only transitions from the v$'$=4 level to a certain low ground state level are expected, for example to the frequency of the transition v$'=4$ --- v$''=0$. This way the selective observation of the desired transitions was already successfully applied for the Ca$_2$ dimer \cite{Allard_2003}. 

The transition path for reaching asymptotic levels from v$''$=0 or 2 via v$'$=4 can be inverted to produce ultracold molecules in v$''$=0 from photoassociation followed by spontaneous decay for populating the asymptotic level.

\section{Acknowledgments}

The authors are grateful to M. Aubert-Fr\'econ and M. Kro\'snicki for providing them with ab initio data about the molecule and M. Yasuda and H. Katori for their partially reproduced scattering wave function of the isotopic combination $^{88}$Sr+$^{88}$Sr used for comparison. The authors also thank G. Gerber for a copy of the Diploma thesis from H. Schneider. 
This work was supported by the Deutsche Forschungsgemeinschaft in
the Sonderforschungsbereich 407.
\bibliography{Sr2_AX_arxiv}

\end{document}